\documentclass[fleqn,12pt,twoside]{article}
\usepackage{espcrc1}
\usepackage{graphicx}

\title{A Heavy Quark Symmetry Approach to Baryons}

\author{C. Albertus\address[dfm]{Departamento de F\'\i sica Moderna.
    Facultad de Ciencias, Universidad de Granada, E-18071 Granada,
    Spain}, J. E. Amaro\addressmark[dfm], E.
  Hern\'andez\address{Grupo de F\'\i sica Nuclear.  Facultad de
    Ciencias, Universidad de Salamanca, E-37008 Salamanca, Spain} and
  J. Nieves\addressmark[dfm]\thanks{ This research was supported by
    DGI and FEDER funds, under contracts BFM2002-03218 and
    BFM2003-00856, by the Junta de Andaluc\'\i a and Junta de Castilla
    y Le\'on under contracts FQM0225 and SA104/04, and it is part of
    the EU integrated infrastructure initiative Hadron Physics Project
    under contract number RII3-CT-2004-506078.  C. Albertus wishes to
    acknowledge a grant related to his Ph.D from Junta de Andaluc\'\i
    a.}}

\begin{document}

\maketitle 

\begin{abstract}
  We evaluate different properties of baryons with a heavy $c$ or $b$ quark.
  The use of Heavy Quark Symmetry (HQS) provides with an important
  simplification of the non relativistic three body problem which  
  can be solved by means of a
  simple variational approach. This scheme is able to reproduce
  previous results obtained with more involved Faddeev calculations.
  The resulting wave functions are parametrized in a simple manner,
  and can be used to calculate further observables.
\end{abstract}

\section{INTRODUCTION}

Since the discovery of $\Lambda_b$ \cite{al91}, \cite{ba91} and most
of the charmed baryons of the SU(3) multiplet on the second level of
the SU(4) 20-plet \cite{pdg02}, a great deal of  theoretical  work 
has been devoted to their study (See for instance 
Refs.~\cite{bow98}-\cite{neu94}).

In this context HQS has proved to be a useful
tool to understand bottom and charmed physics, being
one of the basis of lattice simulations of bottom systems.  HQS is an
approximate SU($N_F$) symmetry of QCD, being $N_F$ the number of heavy
flavours. This symmetry appears in systems containing heavy quarks,
with masses much larger than any other energy scale ($q=\Lambda_{QCD},
m_u, m_d, m_s, \ldots$) controlling the dynamics of the remaining
degrees of freedom. For baryons containing a heavy quark,
 and up to corrections of order
${\cal O}(\frac{q}{m_h})\,\,$\footnote{Here $q$ stands for a typical
  energy scale relevant for the light degrees of freedom while $m_h$
  is the mass of the heavy quark}, HQS guarantees that the heavy
baryon light degrees of freedom quantum numbers are always well defined.
%(see Table~1 of Ref.~\cite{bow98}).

However, HQS has not been systematically used within the context of
non relativistic constituent quark models (NRCQM). The model we present
 here solves the non relativistic three body
problem, for the ground state of baryons with a heavy $c$ or $b$
quark, making full use of the simplifications of HQS \cite{aahn04}. Thanks to HQS, the method proposed provides us with simple wave
functions, while the results obtained for the spectrum and other
observables compare quite well with more sophisticated Faddeev
calculations done in Ref.~\cite{si96}.

\section{THE MODEL}

Once the centre of mass (CM) motion has been removed, the intrinsic
hamiltonian that describes the dynamics of the baryon is
given by%\footnote{In this hamiltonian the motion of the centre of mass
%  (CM) of the baryon has been removed.}
%
\begin{eqnarray}
H^{int}&=&\sum_{i=q,q'}h_i^{sp}+V_{qq'}(\vec{r}_1-\vec{r}_2,spin)-
\frac{\vec{\nabla}_1\cdot\vec{\nabla}_2}{m_Q}+\sum_{i=Q,q,q'}m_i 
\label{eq:ham1}\\
h_{i}^{sp}&=&-\frac{\vec{\nabla}_i^2}{2\mu_i}+V_{Qi}(\vec{r}_i,spin),
\hspace{5mm}i=q,q'
\label{eq:ham2}
\end{eqnarray}
where $\vec r_i$ is the position of the $i$-th light quark ($q,q'$)
with respect to the heavy one ($Q$), $\mu_i$ accounts for the reduced
mass of the heavy and the $i$-th light quark system, $V_{Q\,(q,q')}$
and $V_{q\,q'}$ are the light--heavy and light--light interaction
potentials, and $spin$ stands for possible spin dependence of the
potentials. Note the presence of the Hughes-Eckart term
$\vec\nabla_1\cdot\vec\nabla_2/m_{Q}$ that results from the separation
of the CM motion.

The phenomenological potentials used in this work are the one proposed
in Ref.~\cite{bha81} and the set of potentials introduced in
Ref.~\cite{si96}. We have also considered a potential derived in the
context of the SU(2) linear sigma model in Ref.~\cite{fe93} and that
contains a pattern of spontaneous chiral symmetry breaking.

For the interactions considered, both the total spin  
and the total orbital angular momentum with respect to the heavy quark %,
commute with the intrinsic hamiltonian. Assuming now that the ground
state of the baryons are in $s$-wave, $L=0$, the spatial wave function
can only depend on the relative distances $r_1$, $r_2$ and
$r_{12}=\left|\vec{r}_1-\vec{r}_2\right|$. If we consider the case in
which the heavy quark mass goes to infinity ($m_Q\rightarrow \infty$),
the total spin of the light degrees of freedom also commutes with the
hamiltonian, since the terms of the type
$\vec{\sigma}_Q\cdot\vec{\sigma}_i/m_Q m_i$, $i=q,q'$ vanish. In that 
limiting case the total spin of the light degrees of
freedom is well defined and one can easily write the wave function for
the system (see Ref.~\cite{aahn04} for details).

Even in this limit, solving the three body problem is a nontrivial
task. To do so we adopt a variational approach with a family of
spatial wave functions of the type
\begin{equation}
\Psi_{qq'}^{B_Q}(r_1,r_2,r_{12})=NF^{B_Q}(r_{12})\phi_{q}^{Q}(r_1)\phi_{q'}^{Q}(r_2)
\end{equation}
where $N$ is a normalization constant, $\phi_i^Q$ is the $s$-wave
ground state solution ($\psi_{i}^Q$) of the single particle
hamiltonian ($h_i^{sp}$) corrected at large distances in the form
\begin{equation}
\phi_{i}^Q(r_i)=(1+\alpha_ir_i)\,\psi_{i}^Q(r_i),\hspace{2mm} i=q,q'
\end{equation}
and finally $F^{B_Q}$ is a Jastrow correlation function in the
relative distance of the two light quarks for which we take
\begin{eqnarray}
F^{B_Q}(r_{12}) &=& f^{B_Q}(r_{12})
\sum_{j=1}^4 a_j e^{-b_j^2(r_{12}+d_j)^2},\quad a_1=1 \label{eq:fbig12}\\ 
&&\nonumber \\
f^{B_Q}(r_{12}) &=& \left\{ 
\begin{array}{lcl}  1 -
e^{-cr_{12}} &{\rm if} & V^B_{qq^\prime} (r_{12}=0) \gg 0 \\
&&\\ 1~~ (c\to + \infty) &{\rm if} & V^B_{qq^\prime}
(r_{12}=0) \le 0
\end{array}\right.\label{eq:f12}
\end{eqnarray}
being $\alpha_i$, $a_{i\ne 1}$, $b_i$ and $d_i$ are free variational
parameters.

\section{RESULTS AND CONCLUSIONS}

In this work we have considered the $\Lambda_{b,c}$, $\Sigma_{b,c}$,
$\Xi_{b,c}$ and $\Omega_{b,c}$ baryons, and also the
$\Sigma^*_{b,c}$, $\Xi^*_{b,c}$, $\Xi'_{b,c}$ and $\Omega^*_{b,c}$ baryons
which were not evaluated in Ref.~\cite{si96}. 

Our variational results for charm and bottom masses for the AL1
potential of Ref.~\cite{si96} can be found in Table~\ref{tab:res-al1}.
The results are in good agreement with previous Faddeev calculations
done in Ref.~\cite{si96}\footnote{ Our model does not take into
  account three body terms considered in Ref.~\cite{si96}. Thus, we
  have substracted their effect from the Faddeev results}. They also
agree with lattice results of Ref.~\cite{bow98} and with the
experimental masses.
\begin{table}
\caption {Variational results of charmed and bottom baryons masses (in MeV). 
We also show the Faddeev results of Ref.~\cite{si96},
the lattice results of Ref.~\cite{bow98} and, when
available, the experimental masses~\cite{pdg02}.
$s^\pi$ stands for the spin-parity of the light degrees of freedom.}
\begin{tabular}{cc|cccc|cccc}\hline
  &  &\multicolumn{4}{|c|}{$Q=c$} &\multicolumn{4}{c}{$Q=b$} \\\hline
B            &$s^\pi$ & $M_{exp.}$ & $M_{Latt.}$  & $M_{Var}$ & $M_{Fad.}$
& $M_{exp.}$ & $M_{Latt.}$  & $M_{Var}$ & $M_{Fad.}$ \\\hline 
$\Lambda_Q$  & $0^+$ & $2285 \pm 1$ & $2270 \pm 50$ & $2295$ & $2296$
& $ 5624 \pm 9$& $5640 \pm 60$ & $5643$ & $5643$\\ 
$\Sigma_Q$   &$1^+$  & $2452 \pm 1$ & $2460 \pm 80$ & $2469$ & $2466$&
               & $5770 \pm 70$ & $5851$ & $5849$\\
$\Sigma^*_Q$ & $1^+$ & $2518 \pm 2$ & $2440 \pm 70$ & $2548$ &        &
               & $5780 \pm 70$ & $5882$ & \\ 
$\Xi_Q$      & $0^+$  & $2469 \pm 3$ & $2410 \pm 50$ & $2474$ & $2473$ &
              & $5760 \pm 60$ & $5808$ & $5808$\\ 
$\Xi'_Q$     & $1^+$  & $2576 \pm 2$ & $2570 \pm 80$ & $2578$ &
      &         & $5900 \pm 70$ & $5946$ &  \\ 
$\Xi^*_Q$    & $1^+$ & $2646 \pm 2$ & $2550 \pm 80$ & $2655$ &
      &         & $5900 \pm 80$ & $5975$ &  \\
$\Omega_Q$   &$1^+$  & $2698 \pm 3$ & $2680 \pm 70$ & $2681$ & $2678$
       &         & $5990 \pm 70$ & $6033$ & $6035$ \\ 
$\Omega^*_Q$ & $1^+$  &              & $2660 \pm 80$ & $2755$ &
       &        & $6000 \pm 70$ & $6063$ &  \\\hline 
\end{tabular}\label{tab:res-al1}
\end{table}

\begin{table}
\vspace{-0.5cm}
\caption{Mass mean square radii in $fm^2$ for charmed and bottom baryons.
}
\begin{tabular}{c|cc|cc}
\hline
             &\multicolumn{2}{|c|}{$Q=c$}&\multicolumn{2}{c}{$Q=b$}\\\hline 
B            & $<r^2>_{Var}$ & $<r^2>_{Fad.}$& $<r^2>_{Var}$ & $<r^2>_{Fad.}$\\
\hline
$\Lambda_Q$ &   0.106     &  0.104    &   0.045     &    0.045    \\
$\Sigma_Q$  &   0.123     &  0.121    &   0.057     &    0.054    \\
$\Sigma^*_Q$&   0.135     &           &   0.060     &             \\
$\Xi_Q$     &   0.049     &  0.048    &   0.049     &    0.048    \\
$\Xi'_Q$    &   0.119     &           &   0.060     &             \\
$\Xi^*_Q$   &   0.123     &           &   0.059     &             \\
$\Omega_Q$  &   0.108     &  0.108    &   0.057     &    0.054    \\
$\Omega^*_Q$&   0.120     &           &   0.059     &             \\
\hline
\end{tabular}\label{tab:mradii}
\vspace{-.5cm}
\end{table}

\begin{table}
\caption{Charge mean square radii in $fm^2$ for charmed and bottom baryons.
 We only show results  for baryons with the lesser positive charge.}
\begin{tabular}{c|cc|cc}
\hline
             &\multicolumn{2}{|c|}{$Q=c$}&\multicolumn{2}{c}{$Q=b$}\\\hline 
B            & $<r^2>_{Var}$ & $<r^2>_{Fad.}$& $<r^2>_{Var}$ & $<r^2>_{Fad.}$\\
\hline
$\Lambda_Q$ &     0.131   &  0.129    &   0.127     &    0.128    \\
$\Sigma_Q$  &    -0.261   &  -0.256   &   -0.332    &   -0.318    \\
$\Sigma^*_Q$&    -0.283   &           &   -0.349    &             \\
$\Xi_Q$     &    -0.163   &  -0.161   &   -0.213    &   -0.212    \\
$\Xi'_Q$    &    -0.192   &           &   -0.267    &             \\
$\Xi^*_Q$   &    -0.198   &           &   -0.266    &             \\
$\Omega_Q$  &    -0.124   &  -0.124   &   -0.189    &   -0.183    \\
$\Omega^*_Q$&    -0.138   &           &   -0.196    &             \\
\hline
\end{tabular}\label{tab:cradii}
\vspace{-0.5cm}
\end{table}
Using the wave functions obtained with this method, we have calculated
mass and charge form factors (See Figs.~2-5 of Ref.~\cite{aahn04}),
from which one can obtain mass and charge mean square radii. Our
results for the latter are shown in Tables~\ref{tab:mradii} and
\ref{tab:cradii}. Again we find very good agreement with the results
obtained in Ref.~\cite{si96}. As a further test of the wave functions,
we have also calculated the so called ``wave function at the origin''
(See Ref.~\cite{aahn04}), for which we have good agreement in all
cases, except for the $\Xi$ baryons, with the values obtained in
Ref.~\cite{si96}. The absolute value of this quantity is claimed to be
dependent of the numerical procedure used. Results obtained with the
other interquark interactions can be found in Ref.~\cite{aahn04}.

In this contribution we have outlined the variational scheme developed
 in Ref.~\cite{aahn04} to describe baryons with a
heavy $c$ or $b$ quark. This method for solving the three body problem 
has been possible thanks to the
simplifications introduced by the use of HQS. We have evaluated different
properties of the baryons using several interquark interactions. Our
 results are in good agreement with previous, more involved, Faddeev
calculations done with the same interquark potentials. They also compare well with experimental data and
lattice results. Our wave functions are
much more simpler and manageable than those obtained from the Faddeev
calculation and we have already used them to study
the semileptonic decay of $\Lambda_b$ and $\Xi_b$
baryons \cite{ahnproc04}, \cite{ahn04}

\end{document}